\documentclass[12pt]{article}
\newcommand{\paperonly}[1]{#1}\newcommand{\conferenceonly}[1]{}

\usepackage{epsf,latexsym,amssymb,times}
\usepackage[mathscr]{eucal} %see p.8 of AMSfonts guide; w/o option, redefined
\newcommand{\edo}{\end{document}}

\newcommand{\R}{{\mathbb R}}  %ams bold
\newcommand{\abs}[1]{\left\vert #1 \right\vert}
\newcommand{\norm}[1]{\left\Vert #1 \right\Vert}
\newcommand{\normi}[1]{\left\Vert #1 \right\Vert_\infty }

\newtheorem{theorem}{Theorem}
\newtheorem{itlemma}{Lemma}[section] %number by section (set in \em by default)
\newtheorem{itproposition}[itlemma]{Proposition}
\newtheorem{itcorollary}[itlemma]{Corollary}
\newtheorem{itremark}[itlemma]{Remark}
\newtheorem{itdefinition}[itlemma]{Definition}
\newtheorem{itexample}[itlemma]{Example}

\newenvironment{lemma}{\begin{itlemma}\rm}{\end{itlemma}} %no-italics
\newenvironment{remark}{\begin{itremark}\rm}{\end{itremark}} %no-italics
\newenvironment{corollary}{\begin{itcorollary}\rm}{\end{itcorollary}}
\newenvironment{proposition}{\begin{itproposition}\rm}{\end{itproposition}}
\newenvironment{definition}{\begin{itdefinition}\rm}{\end{itdefinition}}
\newenvironment{example}{\begin{itexample}\rm}{\end{itexample}}

\newcommand{\text}[1]{\hbox{\rm \ #1\ \/}}
\newcommand{\be}[1]{\begin{equation}\label{#1}}
\newcommand{\ee}{\end{equation}}
\newcommand{\bl}[1]{\begin{lemma}\label{#1}}
\newcommand{\ble}[1]{\begin{lemmaex}\label{#1}}
\newcommand{\br}[1]{\begin{remark}\label{#1}}
\newcommand{\bt}[1]{\begin{theorem}\label{#1}}
\newcommand{\bd}[1]{\begin{definition}\label{#1}}
\newcommand{\bp}[1]{\begin{proposition}\label{#1}}
\newcommand{\bc}[1]{\begin{corollary}\label{#1}}
\newcommand{\bfact}[1]{\begin{fact}\label{#1}}
\newcommand{\ber}[1]{\begin{exercise}\label{#1}}
\newcommand{\bex}[1]{\begin{example}\label{#1}}
\newcommand{\bem}[1]{\begin{example}\label{#1}}  %Yes, 2 different ones...
\newcommand{\ec}{\mybox\end{corollary}}
\newcommand{\efact}{\mybox\end{fact}}
\newcommand{\eer}{\mybox\end{exercise}}
\newcommand{\eex}{\mybox\end{example}}
\newcommand{\eem}{\mybox\end{example}}
\newcommand{\el}{\mybox\end{lemma}}
\newcommand{\ele}{\mybox\end{lemmaex}}
\newcommand{\er}{\mybox\end{remark}}
\newcommand{\et}{\qed\end{theorem}}
\newcommand{\ed}{\mybox\end{definition}}
\newcommand{\ep}{\mybox\end{proposition}}
\newcommand{\epr}{\end{proof}}
\newcommand{\bpr}{\begin{proof}}

\newcommand{\ecs}{\end{corollary}}
\newcommand{\eers}{\end{exercise}}
\newcommand{\eexs}{\end{example}}
\newcommand{\eems}{\end{example}}
\newcommand{\els}{\end{lemma}}
\newcommand{\eles}{\end{lemmaex}}
\newcommand{\ers}{\end{remark}}
\newcommand{\ets}{\end{theorem}}
\newcommand{\eds}{\end{definition}}
\newcommand{\eps}{\end{proposition}}
\newcommand{\halmos}{\rule{1ex}{1.4ex}}

\newcommand{\qed}{\hfill \halmos} %put \qed at right margin
\newcommand{\mybox}{\hfill $\Box$} %put \qed at right margin (white square)
\newcommand{\beq}{\begin{eqnarray}}
\newcommand{\eeq}{\end{eqnarray}}
\newcommand{\beqn}{\begin{eqnarray*}}
\newcommand{\eeqn}{\end{eqnarray*}}
\newcommand{\bi}{\begin{itemize}}
\newcommand{\ei}{\end{itemize}}
\newcommand{\ben}{\begin{enumerate}}
\newcommand{\een}{\end{enumerate}}
\newcommand{\st}{\, | \,}

\newenvironment{proof}{\noindent {\em Proof}.\ }{\hspace*{\fill}$\halmos$\medskip}

\newcommand{\bs}{\begin{split}}
\newcommand{\es}{\end{split}}

\newcommand{\Req}{\R_{\geq 0}}

\newcommand{\Rnp}{\Req^n}
\newcommand{\Rmp}{\Req^m}
\newcommand{\Rpp}{\Req^p}
\newcommand{\Rkp}{\Req^k}

\newcommand{\x}{\xi }  % cannot make up my mind if we need a different notation
\newcommand{\z}{\zeta }  % cannot make up my mind if we need a different notation
\newcommand{\uv}{\mu }  

\newcommand{\tilw}{\tilde{w}}
\newcommand{\ip}[2]{\langle #1, #2 \rangle}

\title{Computation of amplification for 
       systems arising from cellular signaling pathways}

\author{Eduardo D.\ Sontag\thanks{Email: {\tt sontag@math.rutgers.edu}.  Supported in part by NIH Grants P20 GM64375 and R01 GM46383.
} \ \ and \ \ 
Madalena Chaves\thanks{Email: {\tt madalena@math.rutgers.edu}.  Supported in part by NIH Grants P20 GM64375 and Aventis.
}\\
{Department of Mathematics,
Rutgers University,
New Brunswick, NJ 08903}}

\begin{document}

\def\today{}
\maketitle

\begin{abstract}
%\noindent
A commonly employed measure of the signal amplification properties of an
input/output system is its induced ${\cal L}^2$ norm, sometimes also
known as $H_\infty $ gain.
In general, however, it is extremely difficult to compute the numerical value
for this norm, or even to check that it is finite, unless the system being
studied is linear.
This paper describes a class of systems for which it is possible to
reduce this computation to that of finding the norm of an associated linear
system. 
In contrast to linearization approaches, a precise value, not an estimate,
is obtained for the full nonlinear model.
The class of systems that we study arose from the modeling of certain
biological intracellular signaling cascades, but the results should be of wider
applicability. 
\end{abstract}
%\begin{keyword}
%norms, gains, nonlinear systems, signaling pathways
%\end{keyword}
%\end{frontmatter}

\section{Introduction}

The analysis of signaling networks constitutes one of the central questions in
systems biology.  There is a pressing need for powerful mathematical tools to
help understand and conceptualize their information processing and
dynamic properties.  One natural question is that of quantifying the amount of
``signal amplification'' in such a network, meaning in some sense the ratio
between the size of a response or output and that of the input that gave rise
to it. See for instance \cite{hnr2002} for a recent paper in this line of work.

In control theory, a routine way to quantify amplification is by means of
the induced ${\cal L}^2$ norm or ``$H_\infty $ gain'' of a system.
A major difficulty when trying to apply these techniques to signaling networks
is that such systems are usually highly nonlinear.
Thus, typically, mathematical results are only given for small inputs
or ``weakly activated'' systems, see for instance \cite{hnr2002,csd2004}.
For large signals, that is, when analyzing the full nonlinear system,
even deciding if the norm is finite or not is usually a very hard question.

In this paper, motivated by the particular systems studied in
\cite{hnr2002,csd2004}, we introduce a class of nonlinear systems, which
includes all these motivational examples as well as many others, and
we show finiteness and how to obtain precise values for norms, by reducing the
problem of norm estimation to the same problem for an associated linear system.
This associated system is sometimes a linearization of the original system
around an equilibrium point, though it need not be.
In any case, the techniques are not at all related to linearization
techniques, but instead borrow from comparison theorems, ISS-like estimates,
and the theory of positive systems.

%We wish to emphasize that positivity of input and state variables plays a key
%role in the theory developed here.  For example, a bilinear system such as
%$\dot x=-(1+u)x + u$ is of the form treated in this paper (just let $A(x)=-1$
%and $B(x)=(1-x)$), and so has finite norm (easily shown to be $1$).
%But if arbitrary, not necessarily nonnegative inputs and states are allowed,
%then the associated i/o operator does not have finite norm.
%%%% oops, not true for L2.. only obvious for ISS.  Think more about it.

\section{Definitions and Statements of Results}

We deal with systems of the following special form:
\be{sys1}
%\dot x=f(x,u)
\dot x(t) = A(x(t))\,x(t) \,+ \,B(x(t))\,u(t)\,,\; x(0)=0
\ee
(or just ``$\dot x=A(x)x+B(x)u$''), where dot indicates time derivative, and
states $x(t)$ as well as input values $u(t)$ are vectors with nonnegative
components: $x(t)\in \Rnp$ and $u(t)\in \Rmp$ for all $t\geq 0$, for some positive
integers $n$ and $m$.  
%%%%%%%%%%%%%%%%%%%%%%%%%%%%%%%%%%%%%%%%%%%%%%%%%%%%%%%%%%%%%%%%%%%%%%%%%%%
We view $A$ and $B$ as matrix valued functions
\[
A: \Rnp \rightarrow  \R^{n\times n}\,,\quad
B: \Rnp \rightarrow  \R^{n\times m}
\]
%%%%%%%%%%%%%%%%%%%%%%%%%%%%%%%%%%%%%%%%%%%%%%%%%%%%%%%%%%%%%%%%%%%%%%%%%%%
where $\Rkp=(\Req)^k$, for any positive integer $k$, is the set of vectors
$\x\in \R^k$ in Euclidean $k$-space with all coordinates $\x_i\geq 0$, $i=1,\ldots ,k$.
%, and similarly $\Req^{k\times \ell}=(\Req)^{k\times \ell}$.
%%%%%%%%%%%%%%%%%%%%%%%%%%%%%%%%%%%%%%%%%%%%%%%%%%%%%%%%%%%%%%%%%%%%%%%%%%%
Associated to these systems we also have an output or measurement
\[
y(t) = h(x(t)) = C(x(t))x(t)
\]
taking values $y(t)\in \R^p$, for some integer $p$, where
$C: \Rnp \rightarrow  \R^{p\times n}$.
%%%%%%%%%%%%%%%%%%%%%%%%%%%%%%%%%%%%%%%%%%%%%%%%%%%%%%%%%%%%%%%%%%%%%%%%%%%
%(Typically, $C$ simply picks out one of the state variables, e.g.\ $Cx=x_n$,
%as a partial read-out of the state of the system.)
%%%%%%%%%%%%%%%%%%%%%%%%%%%%%%%%%%%%%%%%%%%%%%%%%%%%%%%%%%%%%%%%%%%%%%%%%%%

\subsubsection*{Assumptions}

We make several assumptions concerning the matrix functions $A$, $B$, and $C$,
as follows.

\noindent{\em Stability\/}:

The matrix $A(0)$ is Hurwitz, that is, all eigenvalues of $A(0)$ have
negative real parts.

\noindent{\em Maximization at $\x=0$\/}:

For each $\x\in \Rnp$, $A(\x)\leq A(0)$, $B(\x)\leq B(0)$, and
$C(\x)\leq C(0)$, meaning that
$A(\x)_{ij}\leq A(0)_{ij}$ for each $i,j\in \{1,\ldots ,n\}$,
$B(\x)_{ij}\leq B(0)_{ij}$ for each $i\in \{1,\ldots ,n\}$ and $j\in \{1,\ldots ,m\}$.
and
$C(\x)_{ij}\leq C(0)_{ij}$ for each $i\in \{1,\ldots ,p\}$ and $j\in \{1,\ldots ,m\}$.

\noindent{\em Positivity of system\/}:

For each $\x\in \Rnp$ and each $i\in \{1,\ldots ,n\}$ such that $\x_i=0$,
it holds that: $A(\x)_{ij}\geq 0$ for all $j\not= i$ and 
$B(\x)_{ij}\geq 0$ for all $j$.
Also, for every $\x\in \Rnp$, $C_{ij}(\x)\geq 0$ for all $i,j$.

\noindent{\em Local Lipschitz assumption\/}:

The matrix functions $A(\x)$, $B(\x)$, and $C(\x)$ are locally Lipschitz in
$\x$. 

\subsubsection*{Remarks about the form of the system}

The special form assumed for the system is in itself not very restrictive,
since every (affine in controls) system $\dot x=F(x)+B(x)u$ may be written in
this fashion, provided only that $F$ be a continuously differentiable vector
field and $F(0)=0$, for instance by taking
$A(\x) = \int_0^1 F'(\lambda \x)\,d\lambda $, where $F'$ indicates the Jacobian of $F$.  
This reduction to a ``state dependent linear form'' $\dot x=A(x)x+B(x)u$ is
often useful in control theory, where it appears for instance in the context
of ``state-dependent Riccati equation'' approaches to optimal control.
%%%%%%%%%%%%%%%%%%%%%%%%%%%%%%%%%%%%%%%%%%%%%%%%%%%%%%%%%%%%%%%%%%%%%%%%%
%Cloutier, J.R.,
%``State-dependent Riccati equation techniques: an overview,''
%in {\em Proc.\ 1997 American Control Conference\/}, pp. 932-936.
%%%%%%%%%%%%%%%%%%%%%%%%%%%%%%%%%%%%%%%%%%%%%%%%%%%%%%%%%%%%%%%%%%%%%%%%%
Of course, the difficulty is in satisfying the above assumptions for $A$ and
$B$.

A special case in which these hypotheses are satisfied is that of models of
cell signaling cascades as in~\cite{hnr2002,csd2004}.  These are systems whose
equations can be written as follows (with $n$ arbitrary and $m=1$):
\beqn
   \dot x_1 &=& \alpha _1 u (c_1-x_1) -\beta _1x_1  \\
   \dot x_i &=& \alpha _i x_{i-1}(c_i-x_i) -\beta _ix_i\,,  \ \ \ i=2,\ldots ,n
\eeqn
and output $y=x_n$, and
the $\alpha _i$'s, $\beta _i$'s, and $c_i$'s are all positive constants.
We represent this system in the above form using:
\conferenceonly{
$A(\x)_{1,1}=-\beta _1$,
$A(\x)_{i,i-1}=\alpha _ic_i$ for $i=2,\ldots ,n$,
$A(\x)_{i,i}=-\alpha _i\x_{i-1}-\beta _i$ for $i=2,\ldots ,n$,
$B(\x)_{1,1}=\alpha _1c_1-\alpha _1\x_1$,
and all other entries zero.}
\paperonly{
\[
A(\x) \;=\;
\pmatrix{
-\beta _1  &0             & 0             &0                 & \ldots  & 0\cr
\alpha _2c_2&-\alpha _2\x_1-\beta _2& 0             &0                 & \ldots  & 0\cr
0   &  \alpha _3c_3        &-\alpha _3\x_2-\beta _3 &0                 & \ldots  & 0\cr
0   &     0           &\alpha _4c_4        &-\alpha _4\x_3-\beta _4    & \ldots  & 0\cr
0   &  \vdots         &\vdots         &\vdots& \ldots  & \vdots \cr
0   &0                & 0             &0     & \ldots  & -\alpha _n\x_{n-1}-\beta _n
}
\]
and
\[
B(\x) \;=\;
\pmatrix{
\alpha _1c_1-\alpha _1\x_1\cr
0\cr
\vdots\cr
0} \,.
\]
In particular,
\[
A(0) \;=\;
\pmatrix{
-\beta _1    &0              & 0             &0     & \ldots  & 0\cr
 \alpha _2c_2 &-\beta _2          & 0             &0     & \ldots  & 0\cr
     0   &  \alpha _3c_3      &-\beta _3          &0     & \ldots  & 0\cr
     0   &     0         &\alpha _4c_4        &-\beta _4 & \ldots  & 0\cr
     0   &  \vdots       &\vdots         &\vdots& \ldots  & \vdots\cr
     0   &0              & 0             &0     & \ldots  & -\beta _n
}
\]
and
\[
B(0) \;=\;
\pmatrix{
\alpha _1c_1\cr
0\cr
\vdots\cr
0} \,.
\]
}%end paperonly
Note that $A(\x)\leq A(0)$ and $B(\x)\leq B(0)$, for all
$\x\in \Rnp$, because $-\alpha _i\x_i\leq 0$ for all $i$.
The matrix $A(0)$ is lower triangular with negative diagonals, and hence is
Hurwitz.
Positivity holds as well: if $i=1$ and $\x$ is such that $\x_1=0$,
then $A(\x)_{1j}=0$ for all $j\not= 1$ and $B(\x)_{11}=\alpha _1c_1>0$;
if instead $i>1$ and $\x$ is such that $\x_i=0$,
then $A(\x)_{ij}=0$ for all $j\not\in \{i-1,i\}$,
$A(\x)_{i,i-1}=\alpha _ic_i>0$, and $B(\x)_{i1}=0$.
Finally, the functions $A(\cdot )$ and $B(\cdot )$ are linear, and hence Lipschitz.
The matrix $C(\x)=(0,0,\ldots ,0,1)^{\mbox{\sc t}}$ is constant and nonnegative.
Thus all properties hold for this example.

A linear one-dimensional system $\dot x_{n+1}=x_n - \ell x_{n+1}$ may be cascaded
at the end, as in \cite{csd2004}, and the output is in that case redefined as
$y=x_{n+1}$; this may be again modeled in the same way, and the assumptions
still hold. 

\subsubsection*{Induced gains}

Assume given a system~(\ref{sys1}).
We consider the operator $T$ that assigns the
solution function $x$ to each input $u$.  To be more precise, we consider
inputs $u\in {\cal L}^2([0,\infty ),\Rmp)$, and define $x=Tu$ as the unique solution of the
initial value problem~(\ref{sys1}).  In principle, this solution is only
defined on some maximal interval $[0,{\cal T})$, where ${\cal T}>0$ depends on $u$;
however, we will show below that ${\cal T}=+\infty $, and that $x$ is again square
integrable (and nonnegative), so we may view $x$ as an element of
${\cal L}^2([0,\infty ),\Rnp)$ and $T$ as an (nonlinear) operator
\[
T \,:\; {\cal L}^2([0,\infty ),\Rmp) \rightarrow  {\cal L}^2([0,\infty ),\Rnp) \,.
\]
We will write $\abs{\cdot }$ for Euclidean norm,
and use $\norm{\cdot }$ to denote ${\cal L}^2$ norm:
$\norm{u}^2 = \int_0^\infty \abs{u}^2\,dt$.
For the operator $T$, we consider the usual induced operator norm:
\[
\norm{T}\,:=\;
\sup_{u\not= 0}\frac{\norm{Tu}}{\norm{u}} \,.
\]
We will show that $\norm{T}<\infty $ for the systems that we are considering.
In order to see this, we first consider the linear system
\be{sys2}
\dot z \;=\; A(0)z + B(0)u \,,\quad z(0)=0
\ee
with output $v=\ell(z)=C(0)z$,
and its associated operator
\[
L: {\cal L}^2([0,\infty ),\Rmp) \rightarrow  {\cal L}^2([0,\infty ),\Rnp) : u \mapsto  z \, .
\]
Since $A(0)$ is a Hurwitz matrix, $z(t)$ is defined for all $t\geq 0$, and
$L$ indeed maps ${\cal L}^2$ into ${\cal L}^2$.  Furthermore,
its induced norm $\norm{L}$, the ``$H_\infty $ gain'' of the system with output
$y=z$, is finite;  see for instance \cite{control2}.
(The $H_\infty $ gain is defined for arbitrary-valued
inputs $u\in {\cal L}^2([0,\infty ),\R^m)$; we will remark below, cf.\
Section~\ref{positive-inputs-section}, that the same norm is obtained when only
nonnegative inputs are used in the maximization.) 
Moreover, the ${\cal L}^2\rightarrow {\cal L}^\infty $ (or ``$H_2$'') induced gain is also finite.
Therefore, using $\normi{\cdot }$ to denote supremum norm 
$\normi{z} = \sup_{t\geq 0}\abs{z(t)}$,
we can pick a common constant $c\geq 0$ such that
\paperonly{\be{linear-norm-estimates}}
\conferenceonly{\beq\label{linear-norm-estimates}}
\norm{Lu} \leq  c\norm{u} \;\text{and}\;
\normi{Lu} \leq  c\norm{u}
\conferenceonly{\\ \nonumber}
 \;\text{for all}\;
u\in {\cal L}^2([0,\infty ),\Rmp)
\paperonly{\ee}
\conferenceonly{\eeq}
where $c$ upper bounds both $\norm{L}$\ and $\normi{L}$ (we use $\normi{L}$ for
operators to denote induced ${\cal L}^2\rightarrow {\cal L}^\infty $ norm).

Our object of study are the compositions with the output maps, i.e.\ the
input/output operators:
\beqn
T_o &:& {\cal L}^2([0,\infty ),\Rmp) \rightarrow  {\cal L}^2([0,\infty ),\Rpp) \\
&:& u \mapsto  y = C(x)x = C(Tu)Tu
\eeqn
and
\beqn
L_o  &:& {\cal L}^2([0,\infty ),\Rmp) \rightarrow  {\cal L}^2([0,\infty ),\Rpp) \\
 &:& u \mapsto  v = C(0)z = C(0)Lu
\eeqn
and their corresponding induced norms.  Our main result is as follows:

\bt{main-nonlinear-gains}
The norm of $T_o$ is finite, and $\norm{T_o} = \norm{L_o}$.
\ets

\section{Preliminary Results}

We start our proof by remarking that the solutions of~(\ref{sys1}) remain in
$\Rnp$.  To see this, we need to verify the following property (this is a
standard invariance fact; see for instance \cite{monotone} for a discussion in
a related context):

for each $i=1,\ldots ,n$, each $\x\in \Rnp$ such that $\x_i=0$,
and each $\uv\in \Rmp$,
\[
\left(A(\x)\x + B(\x)\uv\right)_i \; \geq  \; 0 \,.
\]
Since $\x_i=0$, we need to prove that
$\sum_{j\not= i}A(\x)_{ij}\x_j + \sum_{j}B(\x)_{ij}\uv_j$
is nonnegative, but this is implied by the positivity assumption.

Similarly, solutions of~(\ref{sys1}) remain in $\Rnp$, as also
$\left(A(0)\x + B(0)\uv\right)_i \geq  0$
if $\x_i=0$.

The next observation is a key one:

\bl{comparison}
Every solution of (\ref{sys1}), with $u\in {\cal L}^2$, is defined for all $t\geq 0$.
Moreover, for any two solutions $x$ of (\ref{sys1}) and (\ref{sys2}) with the
same input $u$, it holds that $0\leq x_i(t)\leq z_i(t)$ for each coordinate
$i=1,\ldots ,n$ and each $t\geq 0$.
\els

\bpr
We use the following comparison principle for differential equations.
Suppose that $f(t,\x)$ and $g(t,\x)$ are such that $f_i(t,\x) \leq  g_i(t,\x)$
for all $i=1,\ldots ,n$ and all $\x\in \Rnp$, and that we consider the solutions of
$\dot x=f(t,x)$ and $\dot z=g(t,z)$ with the same initial condition
(or, more generally, initial conditions $x(0)\leq z(0)$).
Then, provided that $g$ is quasi-monotone (and suitable regularity conditions
hold, as here), we may conclude that
$x(t)\leq z(t)$ (componentwise) for all $t\geq 0$ for which both solutions are
defined.
See for instance \cite{smith,lak}.
Quasi-monotonicity means that $\partial g_i/ \partial \x_j\geq 0$
for all $i\not= j$.

Let us now take any fixed control and let $f(t,\x) = A(\x)\x+B(\x)u(t)$,
$g(t,\x) = A(0)\x + B(0)u(t)$.
We have that $f(t,\x) \leq  g(t,\x)$ coordinatewise, because
$A(\x)\leq A(0)$ and $B(\x)\leq B(0)$ by assumption.
To see that $g$ is quasi-monotone, one needs to verify that
$A(0)_{ij}\geq 0$ for all $i\not= j$. but this follows from the positivity
assumption on $(A,B)$.  Thus the comparison principle tells us that
$x(t)\leq z(t)$ for all $t\geq 0$ for which the solution $x$ is defined
(the solution $z$ is defined for all $t$, since~(\ref{sys2}) is linear
and $A(0)$ is a Hurwitz matrix).
We already observed that $x$ is bounded below by zero; thus,
the maximal solution $x$ is bounded on any finite interval, and hence it is
indeed defined for all $t$, and the Lemma follows.
\epr

\bc{corollary-gains-nonlinear}
For each $u\in {\cal L}^2$, the solution $Tu$ of~(\ref{sys1}) is in ${\cal L}^2$,
and the operator $T$ has finite norm.  Moreover,
\[
\norm{Tu} \leq  \norm{Lu} \leq  c\norm{u}
\]
and
%\;\text{and}\;\;
\[
\normi{Tu} \leq  \normi{Lu} \leq  c\norm{u}
\]
where $c$ is any constant as in~(\ref{linear-norm-estimates}),
so in particular 
$\norm{T} \leq  \norm{L} \leq  c$
and
$\normi{T} \leq  \normi{L} \leq  c$.
Similarly, the i/o operator $T_o$ also has finite norm, 
$\norm{T_ou} \leq  \norm{L_ou}$ and $\normi{T_ou} \leq  \normi{L_ou}$
for all $u\in {\cal L}^2$, and 
$\norm{T_o} \leq  \norm{L_o}$, $\normi{T_o} \leq  \normi{L_o}$.
\ecs

\bpr
Pick any $u$, and let $x=Tu$ and $z=Lu$.
By the Lemma, $0\leq x_i(t)\leq z_i(t)$ for all $t$, so
\[
\norm{x}^2 = \int_0^\infty  \sum_{i=1}^n x_i(s)^2\,ds   \leq 
 \int_0^\infty  \sum_{i=1}^n z_i(s)^2 \,ds = \norm{z}^2.
\]
So $\norm{Tu}\leq \norm{Lu}\leq c\norm{u}$, and since $u$ was arbitrary
it follows that $\norm{T}\leq \norm{L}$.
Similarly,
\[
\normi{x} = \sup_{t\geq 0}\abs{x(t)} \;\leq \;
\sup_{t\geq 0}\abs{z(t)} = \normi{z}
\]
leads to 
$\normi{Tu}\leq  \normi{Lu}$ and $\normi{T}\leq  \normi{L}$.

The positivity and the maximization properties for $C$ imply that, for each
coordinate $i$ of the outputs $y(t)=C(x(t))x(t)$ and $v(t)=C(0)z(t)$, we have
$0\leq y_i(t) = \sum_{j=1}^n C_{ij}(x(t))x_j(t) \leq  
   \sum_{j=1}^n C_{ij}(0)z_j(t) = v_i(t)$,
so the inequalities for $T_o$ and $L_o$ follow by an analogous reasoning.
\epr

Note that the inequality $\norm{T_o} \leq  \norm{L_o}$ gives the finiteness
statement as well as one-half of the equality in the main theorem.

For any matrix $Q$, we denote by $\abs{Q}$ its induced operator norm
as an operator in Euclidean space, that is, the smallest constant $d$ such
that $\abs{Q\x}\leq d\abs{\x}$ for all $\x$.

\bl{Lips-lemma}
There is a nondecreasing and continuous function $M:\Req\rightarrow \Req$ such that:
%\be{estimate-AB}
\[
\abs{A(\x)-A(0)} \leq  M(\abs{\x})\abs{\x}
\]
%\;\;\text{and}\;\;
\[
\abs{B(\x)-B(0)} \leq  M(\abs{\x})\abs{\x}
\]
\[
\abs{C(\x)-C(0)} \leq  M(\abs{\x})\abs{\x}
\]
for all $\x\in \Rnp$.
\els

\bpr
This is a simple consequence of the local Lipschitz property.
On each ball ${\cal B}(R)=\{\xi\st\abs{\x}\leq R\}$, we pick the smallest common
Lipschitz constant $M_0(R)$ for $A(\cdot )$, $B(\cdot )$, and $C(\cdot )$.
The function $M_0$ is nondecreasing, and hence can be majorized by a continuous
and nondecreasing function $M$.
Since $\x\in {\cal B}(\abs{\xi})$, we have that
$\abs{A(\x)-A(0)}\leq M(\abs{\x})\abs{\x}$, and similarly for $B$ and $C$.
\epr

\bc{Lips-cor}
For each function $x\in {\cal L}^2\bigcap {\cal L}^\infty $:
\[
\norm{A(x(\cdot )) - A(0)} \leq  M(\normi{x})\norm{x}
\]
%\;\;\text{and}\;\;
\[
\norm{B(x(\cdot )) - B(0)} \leq  M(\normi{x})\norm{x}
\]
\[
\norm{C(x(\cdot )) - B(0)} \leq  M(\normi{x})\norm{x}
\]
where $M$ is as in Lemma~\ref{Lips-lemma}.
\ecs

\bpr
We have:
\beqn
\norm{A(x(\cdot )) - A(0)}^2
&=&
\int_0^\infty  \abs{A(x(s))-A(0)}^2\,ds\\
&\leq &
\int_0^\infty  M(\abs{x(s)})^2\abs{x(s)}^2 \,ds\\
&\leq &
\int_0^\infty  M(\normi{x})^2\abs{x(s)}^2 \,ds\\
&=&
M(\normi{x})^2 \int_0^\infty  \abs{x(s)}^2 \,ds\\
&=&
M(\normi{x})^2 \norm{x}^2
\eeqn
and similarly for $B$ and $C$.
\epr

\section{Proof of the Main Result}

Pick any input $u\in {\cal L}^2$ and consider once again the respective solutions
$x=Tu$ and $z=Lu$.
By Corollary~\ref{corollary-gains-nonlinear}, we know that 
both $\norm{x}\leq c\norm{u}$ and $\normi{x}\leq c\norm{u}$.
Therefore, using Corollary~\ref{Lips-cor}, we also have that:
\[
\norm{A(x(\cdot )) - A(0)} \leq  c M(c\norm{u})\norm{u}
\]
%\;\;\text{and}\;\;
\[
\norm{B(x(\cdot )) - B(0)} \leq  c M(c\norm{u})\norm{u}
\]
\[
\norm{C(x(\cdot )) - C(0)} \leq  c M(c\norm{u})\norm{u}
\]
where $M$ is as in Lemma~\ref{Lips-lemma}.
Let $\varphi:\Req\rightarrow \Rnp$ be the function $\varphi(t):=$
\[
%\varphi(t) \,:=\;
\left(A(0) - A(x(t))\right)x(t) \,+\,
\left(B(0)-B(x(t))\right)u(t) \,.
\]
By the Cauchy-Schwartz inequality,
\beqn
\norm{\left(A(x(\cdot ))-A(0)\right)x(\cdot )}
&\leq &
\norm{A(x(\cdot ))-A(0)}\norm{x}\\
&\leq &
c^2 M(c\norm{u})\norm{u}^2
\eeqn
and
\beqn
\norm{\left(B(x(\cdot ))-B(0)\right)u(\cdot )}
&\leq &
\norm{B(x(\cdot ))-B(0)}\norm{u}\\
&\leq &
c M(c\norm{u})\norm{u}^2
\eeqn
from which we conclude that
\[
\norm{\varphi} \leq  \gamma (\norm{u})\norm{u}
\]
with $\gamma (r) = (c^2+c) M(cr)r$, and $\gamma $ is a function of class ${\cal K}$,
i.e.\ continuous, strictly increasing, and with $\gamma (0)=0$.

Consider the difference  $w(t)=z(t)-x(t)$.
% it would have been more elegant to write $w(t)=z(t)-x(t)\geq 0$, but I don't
%               want to start changing everything now  :(
Note that $w(0)=0$.
Evaluating 
$\dot w= [A(0)z + B(0)u] - [A(x)x+B(x)u]$ and rearranging terms,
\[
\dot w(t) = A(0) w(t) + \varphi(t)\,.
\]
Using once again that $A(0)$ is a Hurwitz matrix, we know that, for some
constant $d\geq 0$ which depends only on $A(0)$ and not on the particular input
$u$ being used, $\norm{w}\leq d\norm{\varphi}$.
Therefore, $\norm{w}\leq \gamma (\norm{u})\norm{u}$, after redefining
$\gamma (r):=d\gamma (r)$.

%If we consider instead the outputs $y=T_ou = Cx$ and $v=L_ou=Cz$,
%then $v-y = Cw$, and thus $\norm{v-y}\leq \norm{C}\gamma (\norm{u})\norm{u}$, so
%redefining if necessary $\gamma (r):=\norm{C}\gamma (r)$, we may assume that
%also $\norm{L_ou - T_ou} = \norm{v-y}\leq \gamma (\norm{u})\norm{u}$.

In terms of the outputs 
$y=T_ou = C(x)x$ and $v=L_ou=C(0)z$,
\beqn
\norm{v-y} &=& \norm{C(0)z-C(x(\cdot ))x} \\
&\leq & \norm{C(0)(z-x)} + \norm{(C(0)-C(x(\cdot ))x}\\
&\leq & \abs{C(0)}\norm{z-x} + \norm{C(0)-C(x(\cdot ))}\norm{x}\\
&\leq & \abs{C(0)}\gamma (\norm{u})\norm{u} + c^2M(c\norm{u})\norm{u}^2
\eeqn
and we can again write the last term as $\gamma (\norm{u})\norm{u}$ if
we redefine $\gamma (r):= \abs{C(0)}\gamma (r)+c^2M(cr)r$.

The triangle inequality gives us that
$\norm{Lu} -\norm{Tu} \leq  \norm{Lu - Tu}$
and $\norm{L_ou} -\norm{T_ou} \leq  \norm{L_ou - T_ou}$, and
Corollary~\ref{corollary-gains-nonlinear} gives
$\norm{Tu} \leq  \norm{Lu}$ and $\norm{T_ou} \leq  \norm{L_ou}$, so
we may summarize as follows:

\bp{main-technical}
There is a function $\gamma \in {\cal K}$ such that
\[
0\leq  \norm{Lu} -\norm{Tu} \leq  \gamma (\norm{u})\norm{u}
\]
%\;\;\text{and}\;\;
and
\[
0\leq  \norm{L_ou} -\norm{T_ou} \leq  \gamma (\norm{u})\norm{u}
\]
for any input $u\in {\cal L}^2$.
\ep

To conclude the proof of Theorem~\ref{main-nonlinear-gains},
we must show that $\norm{T_o} \geq  \norm{L_o}$.
Let $g = \norm{L_o}$, and pick a minimizing sequence
$u_n$, $n=1,2,\ldots $ of nonzero inputs in ${\cal L}^2$,
that is,
\[
\lim_{n\rightarrow \infty }\frac{\norm{L_ou_n}}{\norm{u_n}} = g \,.
\]
Pick a sequence of real numbers $\varepsilon _n>0$ such that 
$v_n:=\varepsilon _nu_n\rightarrow 0$ (for example, $\varepsilon _n = (n \norm{u_n})^{-1}$).
Since $L_o$ is a linear operator,
$\norm{L_ov_n} = \varepsilon _n \norm{L_ou_n}$, and since $\norm{v_n}=\varepsilon _n\norm{u_n}$,
also 
${\norm{L_ov_n}}/{\norm{v_n}} = {\norm{L_ou_n}}/{\norm{u_n}}$.
Applying the second inequality in Proposition~\ref{main-technical}:
\[
0\,\leq  \, \frac{\norm{L_ov_n}}{\norm{v_n}} - 
\frac{\norm{T_ov_n}}{\norm{v_n}} \, \leq  \, \gamma (\norm{v_n}) \rightarrow  0
\]
which gives that $\frac{\norm{T_ov_n}}{\norm{v_n}}\rightarrow g$, and therefore
$\norm{T_o}\geq c$, as desired.
\qed

\section{Positive vs.\ arbitrary inputs}
\label{positive-inputs-section}

We have shown that the norm of the nonlinear system~(\ref{sys1}) can be
exactly computed by finding the norm of the associated linear
system~(\ref{sys2}).  The computation of induced ${\cal L}^2$ norms for linear
systems is a classical area of study, and amounts to the maximization, over
the imaginary axis, of the largest singular value of the transfer matrix of
the system (the Laplace transform of the impulse response), the $H_\infty $ norm;
see for instance~\cite{control2}.  There is, however, a potential gap in the
application of this theory to our problem, namely, the usual definition of
$H_\infty $ norm corresponds to maximization over {\em arbitrary\/} inputs
$u\in {\cal L}^2([0,\infty ),\R^m)$, not necessarily inputs with values in $\Rmp$ as
considered in this paper.  We close this gap now, by showing that the same
result is obtained, for systems~(\ref{sys2}), whether one optimizes over
arbitrary or over nonnegative inputs.  We give two proofs, one elementary and
the other one less trivial but leading to a stronger conclusion. 

The positivity assumptions imply that the operator $L_o$ is a nonnegative
convolution operator:
\beq
\label{pos-operator}
(L_ou)(t) &=& \int_0^t W(t-s) u(s) \,ds\,,
%\quad
\\
\label{pos-operator1}
&&\quad W(t)\in (\Req)^{p\times m} \;\forall\,t\geq 0\,.
\eeq
Here $W(t) = C(0)e^{tA(0)}B(0)$, and its nonnegativity follows from the fact that
$e^{tF}$ has all entries nonnegative, provided that $F_{ij}\geq 0$ for all $i\not= j$.
(This last fact is well-known: it is clear for small $t$ from the expansion
$e^{tF} = I+tF+o(t)$, and for large $t$ by then writing $e^{tF}$ as a product
of matrices $e^{(t/k)F}$ with the positive integer $k$ large enough.)
We next show that any operator as in~(\ref{pos-operator}-\ref{pos-operator1})
has the same norm whether viewed as an operator on ${\cal L}^2([0,\infty ),\R^m)$ or on
${\cal L}^2([0,\infty ),\Rmp)$. 
%%%%%%%%%%%%%%%%%%%%%%%%%%%%%%%%%%%%%%%%%%%%%%%%%%%%%%%%%%%%%%%%%%%%%%%%%
Since the norm as an operator on nonnegative inputs is, obviously, upper
bounded by the norm on arbitrary inputs, it will be enough to show that, for
each $w\in {\cal L}^2([0,\infty ),\R^m)$, there is another input $\tilw\in {\cal L}^2([0,\infty ),\Rmp)$
with $\norm{w}=\norm{\tilw}$ and $\norm{L_ow}\leq \norm{L_o\tilw}$.

Given such a $w$, we start by writing $w = u-v$, where $u$ and $v$ are picked
in ${\cal L}^2([0,\infty ),\Rmp)$ and orthogonal.
(Such a decomposition is always possible.
We define coordinatewise, for each $i=1,\ldots ,m$, $u_i := \max\{w_i,0\}$ and
$v_i := \max\{-w_i,0\}$; clearly, $w=u-v$.
The supports of $u_i$ and $v_i$ are disjoint, 
so $\ip{u_i}{v_i}=\int_0^\infty u_i(t)v_i(t)\,dt=0$ for each $i$, and also then
$\ip{u}{v} = \sum_{i=1}^m \ip{u_i}{v_i} = 0$.)
We now let $\tilw:=u+v$.
Since $u$ and $v$ (or $-v$) are orthogonal, 
$\norm{w}^2=\norm{u}^2+\norm{-v}^2=\norm{u}^2+\norm{v}^2=\norm{\tilw}^2$,
so $\norm{w}=\norm{\tilw}$.
Because $L_o$ is nonnegative, both $x=L_ou$ and $y=L_ov$ are nonnegative.
To finish the proof, we only need to see that $\norm{x-y}\leq \norm{x+y}$:
\beqn
\norm{x-y}^2 &=& \int_0^\infty  {\textstyle\sum_{i=1}^p} (x_i(t)-y_i(t))^2\,dt\\
             &=& \int_0^\infty  {\textstyle\sum_{i=1}^p} (x_i(t)^2 + y_i(t)^2 - 2x_i(t)y_i(t))\,dt\\
             &\leq & \int_0^\infty  {\textstyle\sum_{i=1}^p} (x_i(t)^2 + y_i(t)^2 + 2x_i(t)y_i(t))\,dt\\
             &=& \int_0^\infty  {\textstyle\sum_{i=1}^p} (x_i(t)+y_i(t))^2\,dt\\
             &=& \norm{x+y}^2 \,.
\eeqn

A different proof, which in fact also implies that the supremum in the
definition of norm is achieved as a maximum, is as follows.
We consider the adjoint $L_o^*$ of $L_o$ (seen as an operator on the Hilbert
space ${\cal L}^2([0,\infty ),\R^m)$), and the composition 
$M = L_o^*L_o:{\cal L}^2([0,\infty ),\R^m)\rightarrow {\cal L}^2([0,\infty ),\R^m)$.
The operator $M$ is self-adjoint and (since $L_o$ is a convolution operator
with an ${\cal L}^2$ kernel) compact.  Its spectrum consists of real and nonnegative
eigenvalues, and its largest eigenvalue $\lambda $ is such that $\mu =\sqrt{\lambda }$
is the largest singular value of $L_o$, and equals the norm of $L_o$
as an operator ${\cal L}^2([0,\infty ),\R^m)\rightarrow {\cal L}^2([0,\infty ),\R^p)$.
Take any eigenvector $u$ corresponding to $\lambda $, so $Mu=\lambda u$.
It follows that
$\norm{L_ou}^2=\ip{L_ou}{L_ou}=\ip{u}{Mu} = \ip{u}{\lambda u} = \mu ^2\norm{u}^2$,
so $u$ is a maximizing vector for $L_o$.
Moreover, for a compact positive operator $M$ on a Hilbert space, the
Krein-Rutman Theorem says that, provided that there is a nonzero eigenvalue
(which there is in this case, since $M$ is self-adjoint and we may assume
without loss of generality that $M\not= 0$), then the maximal
eigenvalue $\lambda $ admits a nonnegative eigenvector $u$.
Thus $\norm{L_ou}$ is maximized at this $u\in {\cal L}^2([0,\infty ),\Rmp)$.

\section{Cascades}

Signaling systems are often built by cascading subsystems, so it is
interesting to verify that a cascade of any number of systems which satisfy
our properties again has the same form.  It is enough, by induction, to show
this for two cascaded systems
\beqn
\dot x &=& A_1(x)x+B_1(x)u\,\;\; v=C_1(x)x\\
\dot z &=& A_2(z)z+B_2(z)\tilde{u}\,\;\; y=C_2(z)z
\eeqn
each of which satisfies our assumptions, under the series connection obtained
by setting $\tilde{u}=v$.
The composite system can be represented in terms of the following $A(\x,\z)$
and $B(\x,\z)$ matrices:
\[
A = \pmatrix{A_1(\x)        & 0\cr
                 B_2(\z)C_1(\x) & A_2(\z)} \,,\;
B = \pmatrix{B_1(\x)\cr 0}
\]
and output $y$.

It is easy to verify all the necessary properties.
For example, the only nontrivial part of the maximization property amounts to
checking that $B_2(\z)C_1(\x)\leq B_2(0)C_1(0)$, which follows from
$B_2(\z)C_1(\x)\leq B_2(0)C_1(\x)$ (using the maximization property for
$B_2$ and the positivity of $C_1$) and
$B_2(0)C_1(\x)\leq B_2(0)C_1(0)$ (using maximization for $C_1$ and positivity
of $B_2(0)$).
Similarly, the only nontrivial part of the positivity property involves
checking that $(B_2(\z)C_1(\x))_{ij}\geq 0$ provided that $\z_i=0$, for all $j$.
But, for such a vector $\z$, we know that $B_2(\z)_{ik}\geq 0$ for all $k$,
so indeed $\sum_k B_2(\z)_{ik} C_1(\x)_{kj}\geq 0$.

\section{Remarks and Conclusions}
\label{remarks-section}

We provided a way to compute, for systems of a special form, the induced
${\cal L}^2$ norm of the system.
The special form includes a variety of cellular signaling cascade systems.
An even wider class of systems can be included as well, provided that one
extend our treatment to systems that are monotone with respect to orders
other than that given by the first quadrant.  Such orders have proven useful
in analyzing, for example, MAPK cascades, see for example~\cite{monotone,PNAS}.
The details of this extension will be provided elsewhere.

\edo